\documentclass[twocolumn]{aa}  
\usepackage{thesis_defs}
\usepackage{graphicx}   
\usepackage{placeins}
\usepackage{pifont}
\usepackage[colorlinks=true,
linkcolor=blue,
citecolor=blue]{hyperref}
\usepackage{booktabs}
\usepackage{subcaption}
\usepackage{tikz}
\usepackage{upgreek}
\usepackage[export]{adjustbox}
\usepackage{orcidlink}

\usepackage{txfonts}
\usepackage{hyperref}
\makeatletter
\renewcommand\makeLineNumber{}   
\makeatother

\begin{document}

\title{Three-dimensional Simulations and Synthetic Polarization of the SS\,433 Jet}

   \author{Joana A. Kramer \orcidlink{0009-0003-3011-0454}
          \inst{1,2*}
          \and
          Haocheng Zhang 
          \inst{3,4}
          \and
          Hui Li 
          \inst{2,5}
          }

   \institute{ \inst{1}Anton Pannekoek Institute for Astronomy ,University of Amsterdam, Science Park 904, 1098 XH, Amsterdam, NL\\
    \inst{*}\email{j.a.hoellmer@uva.nl}\\
    \inst{2}Theoretical Division, Los Alamos National Laboratory, Los Alamos, NM 87545, USA\\
    \inst{3}University of Maryland Baltimore County, Baltimore, MD 21250, USA\\
    \inst{4}NASA Goddard Space Flight Center, Greenbelt, MD 20771, USA\\
    \inst{5}New Mexico Consortium, Los Alamos, NM 87544, USA
     }

   \date{\today}
   
 \abstract{SS~433 is the prototypical microquasar and the only known Galactic system hosting persistent, precessing relativistic jets powered by supercritical accretion. Recent X-ray polarimetric observations have revealed highly ordered magnetic fields and stationary high-energy emission regions that are likely associated with recollimation shocks.}
{We investigate how magnetic-field topology influences the formation of recollimation shocks, jet morphology, and polarized synchrotron emission in SS~433-like jets, with the goal of constraining the magnetic structure responsible for the observed X-ray polarization.}
{We performed a suite of three-dimensional relativistic magnetohydrodynamic simulations using the PLUTO code, exploring purely poloidal, purely toroidal, and mixed helical magnetic-field configurations. The resulting magnetic and thermodynamic structures were post-processed with the polarized radiative-transfer code 3DPol to generate synthetic synchrotron intensity and polarization maps directly comparable to observations.}
{We find that the magnetic topology is the primary regulator of jet collimation and shock development. Toroidal-field jets produce the narrowest outflows and the strongest sequence of recollimation shocks, whereas poloidal-field jets remain broader and dynamically smoother. The large-scale polarization structure is remarkably insensitive to jet radius and non-thermal particle normalization. Instead, the polarization signatures are governed almost entirely by the magnetic geometry. Poloidal-dominated configurations produce high and stable polarization fractions, supported by the IXPE observations of SS~433, with magnetic fields aligned with the jet direction, whereas toroidal-dominated models generate more complex polarization structures and transverse signatures.}{} 

 \keywords{relativistic magnetohydrodynamics (RMHD) -- numerical -- polarization -- non-thermal -- shock waves -- jets -- SS 433 }
   
  \authorrunning{J.~A.~Kramer et al.}
  \titlerunning{Synthetic Polarization of the SS 433 Jet}
 \maketitle
\section{Introduction}\label{sec:intro}

SS~433, one of the most enigmatic X-ray binaries, has been a subject of significant astrophysical interest since its discovery. This system is unique due to its continuous, precessing relativistic jets, which are launched at approximately 0.26\,c, making it the only known microquasar to exhibit persistent supercritical accretion~\citep{Margon84, Fabrika04}. As the prototype of microquasars, SS~433 provides invaluable insights into jet physics, accretion processes, and supercritical flows, drawing parallels between stellar-mass and supermassive black hole (SMBH) systems~\citep{Blundell&Bowler04, Kaaret24}. The system consists of a compact object, most likely a stellar-mass black hole or neutron star, and a massive donor star, whose mass transfer drives the extreme accretion rates~\citep{Begelman06}. The optical spectrum of SS~433 displays emission lines with Doppler shifts corresponding to the jet velocities, revealing the system’s kinematic and dynamic complexity~\citep{Hillwig08}.

Despite extensive multi-wavelength observations, several key aspects remain uncertain—particularly the nature of the compact object, the mechanisms driving the jet collimation, and how the jets interact with the surrounding nebula W50~\citep{Safi22}. Observations made with the Very Large Array (VLA)~\citep{Hjellming81,Blundell&Bowler04}, NASA's Imaging X-ray Polarimetry Explorer (IXPE)~\citep{Weisskopf22, Kaaret24}, H.E.S.S.~\citep{Hess2024}, LHASSO~\citep{LHAASO}, and HAWC~\citep{Fang20} have revealed the structure and behavior of the jets in radio, X-rays, and gamma rays, respectively. These studies confirm the precession of the inner jets as well as discuss the presence of localized emission features
as they propagate into W50~\citep{Goodall}.

IXPE observations have provided key insights into the synchrotron nature of the X-ray emission in SS\, 433~\citep{Kaaret24}. They reveal stationary features further down the jet ($\sim$ 30\,parsec (pc) outside the BH system) interpreted as recollimation shocks—structures. Those are theoretically predicted as natural outcomes of pressure focusing during jet propagation~\citep{DalyMarscher, Kramer&MacDonald}. In analogy to AGN jets, these recollimation shocks may represent stable acceleration regions where magnetic fields and flow geometry remain steady over long timescales.

A magnetized plasma description is appropriate for the jets, given their large-scale ordered magnetic fields~\citep[confirmed by X-ray electric vector position angle (EVPA) orientation][]{Kaaret24}. Using relativistic magnetohydrodynamics (RMHD), one can model the jets under the continuum approximation when the jet radius $r_\text{jet}$ is much larger than the plasma kinetic scales
~\citep{Hawley2015}. Jet launching is believed to occur via magneto-centrifugal processes from the accretion disk, e.g., Blandford–Payne mechanism~\citep{BlandfordPayne}, possibly in combination with the extraction of rotational energy from the compact object, e.g., 
Blandford–Znajek mechanism~\citep{BlandfordZnajek}. These processes naturally produce helical or toroidal magnetic fields, which can be compared with polarimetric observations~\citep{BlandfordPayne,BlandfordZnajek,Hardee2007,Gabuzda2008,Gabuzda2018}.

Helical magnetic structures can lead to current-driven instabilities and magnetic reconnection events~\citep{Kadowaki}, which produce non-thermal particle populations with power-law energy distributions. Particle-in-cell (PIC) simulations~\cite[e.g.,][]{Sironi2021} confirm that reconnection events can accelerate electrons efficiently, producing non-thermal synchrotron and inverse-Compton radiation up to X-ray and gamma-ray energies.

In hydrodynamic approaches, non-thermal particle (NTP) populations are treated as tracers within the thermal plasma. Shock acceleration at internal discontinuities—arising naturally from precession and intermittent mass ejections—produces power-law particle distributions ($N(E) \propto E^{-s}$). Typical indices ($s = 2.0-2.3$) are found in RMHD studies~\citep{delaCita, Fromm2016, Kramer&MacDonald}, consistent with the measured X-ray photon indices of SS~433~\citep{Kaaret24}. These shocks are thought to correspond to structural features seen in imaging observations, including the X-ray knots where jets interact with the surrounding medium.

The remainder of this paper is organized as follows. In Sec.~\ref{sec:observations}, we summarize the current observational and theoretical understanding of SS~433, with particular emphasis on the recollimation shock, particle acceleration, and polarization properties. Section~\ref{sec:style} describes the numerical methods, including the relativistic magnetohydrodynamic simulations, the adopted magnetic-field configurations, and the polarized radiative-transfer calculations. In Sec.~\ref{sec:mag}, we present the magnetohydrodynamic evolution of the simulated jets and investigate how different magnetic topologies affect the jet morphology, shock structure, and thermodynamic properties. Section~\ref{sec:polarization} focuses on the resulting magnetic flux evolution and synthetic polarized synchrotron emission, highlighting the observational signatures of different magnetic-field configurations. In Sec.~\ref{sec:disc}, we discuss our findings in the context of recent multi-wavelength observations of SS~433 and their implications for jet magnetic structure and particle acceleration. Finally, our main conclusions are summarized in Sec.~\ref{sec:conc}.


\section{SS\,433: observational context and simulation targets}
\label{sec:observations}
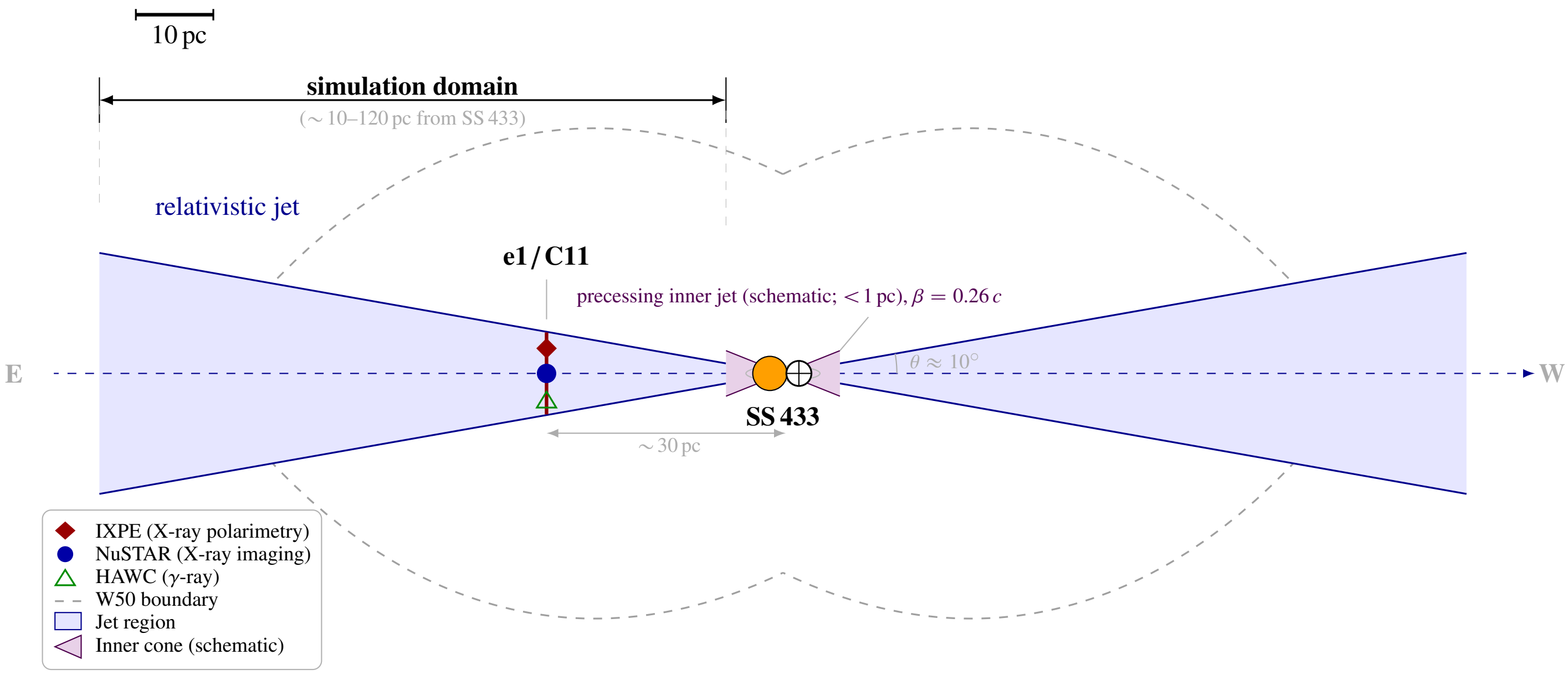
\begin{figure*}
\centering
\begin{tikzpicture}[
  scale  = 0.68,
  >=latex,                          
  every node/.style = {font=\small} 
]

\draw[gray!75, dashed, semithick]
  (0, 3.5)
    .. controls (3.5, 5.2) and (7.0, 4.2) .. (9.8,  0.6)
    .. controls (10.4, 0.2) and (10.4,-0.2) .. (9.8, -0.6)
    .. controls (7.0,-4.2) and (3.5,-5.2) ..
  (0,-3.5);
\draw[gray!75, dashed, semithick]
  (0, 3.5)
    .. controls (-3.5, 5.2) and (-7.0, 4.2) .. (-9.8,  0.6)
    .. controls (-10.4, 0.2) and (-10.4,-0.2) .. (-9.8,-0.6)
    .. controls (-7.0,-4.2) and (-3.5,-5.2) ..
  (0,-3.5);
\node[gray!75, font=\itshape\footnotesize] at ( 11.0, 0) {W50};
\node[gray!75, font=\itshape\footnotesize] at (-11.0, 0) {W50};

\fill[blue!10]
  (1, 0.176) -- (12, 2.116) -- (12,-2.116) -- (1,-0.176) -- cycle;
\fill[blue!10]
  (-1, 0.176) -- (-12, 2.116) -- (-12,-2.116) -- (-1,-0.176) -- cycle;
\draw[blue!55!black, semithick] (1, 0.176) -- (12, 2.116);
\draw[blue!55!black, semithick] (1,-0.176) -- (12,-2.116);
\draw[blue!55!black, semithick] (-1, 0.176) -- (-12, 2.116);
\draw[blue!55!black, semithick] (-1,-0.176) -- (-12,-2.116);
\draw[blue!55!black, dashed, thin, ->] (0, 0) -- (13.2, 0);
\draw[blue!55!black, dashed, thin    ] (0, 0) -- (-12.8, 0);
\node[blue!55!black, font=\footnotesize, anchor=south west] at (-11.2, 2.5)
  {relativistic jet};

\draw[gray!65, thin] (2.0, 0) arc (0:10:2.0);   
\node[gray!65, font=\scriptsize, anchor=west] at (2.05, 0.22)
  {$\theta \approx 10^{\circ}$};

\fill[violet!18!white] (0,0) -- ( 1.0, 0.40) -- ( 1.0,-0.40) -- cycle;
\draw[violet!60!black, thin] (0,0) -- ( 1.0, 0.40)
                              (0,0) -- ( 1.0,-0.40);
\fill[violet!18!white] (0,0) -- (-1.0, 0.40) -- (-1.0,-0.40) -- cycle;
\draw[violet!60!black, thin] (0,0) -- (-1.0, 0.40)
                              (0,0) -- (-1.0,-0.40);
\draw[gray!60, thin] (1.0, 0.40) -- (1.5, 0.99);
\node[violet!65!black, font=\scriptsize, anchor=south west] at (-3.8, 0.99)
  {precessing inner jet (schematic; $<\!1\,$pc), $\beta=0.26\,c$};

\draw[gray!55, thin] (0,0) ellipse (0.65 and 0.13);
\fill[orange!75!yellow] (-0.23, 0) circle (0.30);
\draw[black, thin      ] (-0.23, 0) circle (0.30);
\fill[white            ] ( 0.28, 0) circle (0.22);
\draw[black, semithick ] ( 0.28, 0) circle (0.22);
\draw[black, thin      ] ( 0.07, 0) -- (0.49, 0);
\draw[black, thin      ] ( 0.28,-0.22) -- (0.28, 0.22);
\node[font=\small\bfseries, anchor=south] at (0, -1.1) {SS\,433};

\draw[red!55!black, very thick] (-4.15,-0.72) -- (-4.15, 0.72);

\fill[red!60!black]
  (-4.15, 0.62) -- (-4.33, 0.44) -- (-4.15, 0.26) -- (-3.97, 0.44) -- cycle;

\fill[blue!65!black] (-4.15,0.0) circle (0.17);

\draw[green!55!black, semithick]
  (-3.98,-0.6) -- (-4.15,-0.32) -- (-4.32,-0.60) -- cycle;

\draw[gray!60, thin] (-4.15, 0.95) -- (-4.15, 1.65);
\node[font=\small\bfseries, anchor=south] at (-4.15, 1.65) {e1\,/\,C11};

\draw[<->, gray!65, thin] (0.05,-1.05) -- (-4.15,-1.05);
\node[gray!65, font=\scriptsize] at (-2.0,-1.30) {$\sim\!30\,$pc};

\draw[<->, semithick] (-1,4.8) -- (-12,4.8);
\draw[thin] ( -1,4.4) -- ( -1,5.2);
\draw[thin] (-12,4.4) -- (-12,5.2);
\draw[gray!55, dashed, very thin] ( -1,2.6) -- ( -1,4.7);
\draw[gray!55, dashed, very thin] (-12,3.0) -- (-12,4.7);
\node[font=\small\bfseries, anchor=north] at (-6.5,5.4)
  {simulation domain};
\node[gray!65, font=\scriptsize, anchor=north] at (-6.5,4.8)
  {($\sim\!10$--$120\,$pc from SS\,433)};

\draw[very thick] (-10.0,6.3) -- (-11.36,6.3);
\draw[semithick]  (-10.0,6.2) -- (-10.0,6.4);
\draw[semithick]  (-11.36,6.2) -- (-11.36,6.4);
\node[font=\footnotesize, anchor=north] at (-10.6,6.3) {10\,pc};


\node[gray!65, font=\footnotesize\bfseries] at ( 13.5, 0) {W};
\node[gray!65, font=\footnotesize\bfseries] at (-13.5, 0) {E};

\begin{scope}[shift={(-12.8,-1.8)}]
  \draw[gray!60, thin, rounded corners=3pt]
    (-0.2, -0.6) rectangle (4.7,-3.4);
  \fill[red!60!black]
    (0.20, -1.12) -- (0.38, -0.96) -- (0.20, -0.80) -- (0.02, -0.96) -- cycle;
  \node[anchor=west, font=\scriptsize] at (0.55, -1.0)
    {IXPE (X-ray polarimetry)};
  \fill[blue!65!black] (0.20,-1.375) circle (0.14);
  \node[anchor=west, font=\scriptsize] at (0.55,-1.4)
    {NuSTAR (X-ray imaging)};
  \draw[green!55!black, semithick]
    (0.03,-1.92) -- (0.20,-1.65) -- (0.37,-1.92) -- cycle;
  \node[anchor=west, font=\scriptsize] at (0.55,-1.8)
    {HAWC ($\gamma$-ray)};
  \draw[gray!75, dashed, semithick] (0.02,-2.2) -- (0.48,-2.2);
  \node[anchor=west, font=\scriptsize] at (0.55,-2.2) {W50 boundary};
  \fill[blue!10] (0.02,-2.4) rectangle (0.48,-2.7);
  \draw[blue!55!black, thin] (0.02,-2.4) rectangle (0.48,-2.7);
  \node[anchor=west, font=\scriptsize] at (0.55,-2.6) {Jet region};
  \fill[violet!18!white]
    (0.02,-3.0) -- (0.48,-2.8) -- (0.48,-3.2) -- cycle;
  \draw[violet!60!black, thin]
    (0.02,-3.0) -- (0.48,-2.8) -- (0.48,-3.2) -- cycle;
  \node[anchor=west, font=\scriptsize] at (0.55,-3.0)
    {Inner cone (schematic)};
\end{scope}

\end{tikzpicture}
\caption{Schematic of the SS\,433\,/\,W50 system (not to scale).
The central binary 
launches bipolar relativistic jets ($\beta = 0.26\,c$) that precess with a
period of ${\sim}162$\,d.
At parsec scales the precessing inner jet (schematic wedge; extent exaggerated,
true size $<\!1$\,pc) time-averages into a collimated outflow with half-opening
angle $\theta \approx 10^{\circ}$.
The W50 radio nebula (grey dashed contour) is elongated along the jet axis by the
jet--ISM interaction, with ears extending to ${\sim}100$\,pc from the binary.
Coloured symbols mark the quasi-stationary X-ray and $\gamma$-ray emission knots
detected by IXPE (diamond), NuSTAR (circle), and HAWC (triangle) at
${\sim}25$--$35$\,pc from the binary (eastern knots e1\,/\,C11), identified as
recollimation shocks.
The double-headed arrow indicates the simulation domain analysed in this work
(${\sim}10$--$120$\,pc from SS\,433).
}
\label{fig:sketch}
\end{figure*}

SS\,433 is a high-mass X-ray binary at a distance of $5.5\,\mathrm{kpc}$
\citep{Hjellming81}, embedded in the radio nebula W50.
The system launches persistent, bipolar relativistic jets at
$\beta \approx 0.26\,c$ \citep{Margon84,Fabrika04} that precess with
a period of $\sim 162\,\mathrm{days}$ around an axis inclined at
$\approx 20^{\circ}$ to the plane of the sky \citep{Hillwig08}.
Figure~\ref{fig:sketch} illustrates the multi-scale architecture of the
system: the inner precessing jets ($\lesssim 1\,\mathrm{pc}$) inflate a
pair of large-scale lobes that punch through the W50 shell out to
projected distances of $\sim 100\,\mathrm{pc}$ on each side of the
binary.
A recent study by \citet{Gasealahwe25} provides compelling new evidence that SS \,33 is not an isolated phenomenon
but rather part of a broader class of young jet–SNR systems. Their study motivates us to believe that we can draw a strong parallel to
SS\,433 and its surrounding nebula W50, suggesting that Cir X-1
may represent a younger analogue of such jet-SNR systems.
It is noteworthy that we do not attempt to reproduce the inner precessing jet or the detailed
interaction between the jet and the W50 shell.

\subsection{Multi-wavelength structure of the large-scale jet}

Radio observations with the Very Large Array \citep[VLA;][]{Hjellming81,Blundell&Bowler04} resolved the bipolar morphology of SS\,433 and established
its orientation: the eastern and western jet axes are inclined at position
angles of approximately $100^{\circ}$ and approximately $80^{\circ}$ relative to the
line of sight, respectively.
At projected distances of $\sim 30\,\mathrm{pc}$ from the binary, the jets
display a non-thermal radio-to-X-ray spectrum, with localized X-ray knots
detected by XMM-Newton and NuSTAR in the $3$--$30\,\mathrm{keV}$ band
\citep{Safi97,Safi99,Safi22}.
The X-ray spectra show a power-law photon index $\Gamma = 1.6$,
consistent with synchrotron emission from shock-accelerated electrons.
Strikingly, these knots lack counterparts at other wavelengths on the
same spatial scales, suggesting compact, quasi-stationary acceleration
regions rather than extended thermal emission
\citep{Safi22,Kaaret24}.
High-energy $\gamma$-ray observations have confirmed that the jets of
SS\,433 accelerate particles to extreme energies.
HAWC detected $\sim$TeV emission that is spatially coincident with the
X-ray knots \citep{Abeysekara18}, a result subsequently confirmed and
extended to hundreds of TeV by H.E.S.S.\ \citep{Hess2024}
and LHAASO \citep{LHAASO}.
The co-spatiality of the X-ray synchrotron and TeV $\gamma$-ray peaks
strongly argues for \emph{in-situ} particle acceleration at standing
internal shocks in the jet, rather than in the extended jet--W50
interaction zone \citep{Abeysekara18,Goodall,Hess2024}.
The $\gamma$-ray spectral energy distribution is best reproduced by a
leptonic model \citep[inverse-Compton scattering of CMB and local radiation
fields by electrons with $E_e \lesssim 130\,\mathrm{TeV}$;][]{Hess2024},
although hadronic contributions from relativistic protons
($E_p \sim 250\,\mathrm{TeV}$) colliding with ambient material cannot be
excluded \citep{Fang20}.

\subsection{X-ray Polarimetry and the recollimation shock interpretation}

The Imaging X-ray Polarimetry Explorer \citep[IXPE;][]{Weisskopf22}
has provided the most direct window yet onto the magnetic structure of the
SS\,433 jets \citep{Kaaret24}.
IXPE measured a polarization degree of $62\% \pm 15\%$ in the eastern
lobe and $59\% \pm 17\%$ in the C11 region in the $2$--$8\,\mathrm{keV}$
band, with an electric-vector position angle (EVPA) of
$-12^{\circ} \pm 10^{\circ}$ that is approximately constant across both
regions.
For synchrotron emission, the EVPA is perpendicular to the local magnetic
field, so the measured EVPA implies that the magnetic field is aligned
\emph{parallel} to the jet axis throughout the recollimation zone.
The high polarization fraction — exceeding 50\% — implies that at least
half the magnetic field is spatially ordered at the resolution of IXPE,
ruling out a predominantly turbulent field geometry.
The associated X-ray luminosity is
$L \approx 1.3 \times 10^{35}\,\mathrm{erg\,s^{-1}}$ in the
$0.3$--$30\,\mathrm{keV}$ band \citep{Kaaret24}.
These observations are naturally interpreted in the framework of
recollimation shocks \citep{DalyMarscher,Kramer&MacDonald}: quasi-stationary oblique
shocks that form where the overpressured jet recollimates against the
confining ambient medium at distances of $\sim 25$--$30\,\mathrm{pc}$
from the binary.
Compression and shear at such shocks preferentially align the magnetic
field with the jet axis \citep{Giacalone07}, boosting the polarization
fraction and fixing the EVPA, consistent with the IXPE measurements.
The stationarity of the X-ray knots over decade-long baselines
\citep{Kaaret24} supports a standing-shock interpretation rather than
moving ejecta or the advancing jet head.
Magnetic-field estimates inferred from the synchrotron spectrum place
$B \approx 10$--$20\,\mu\mathrm{G}$ in the emission region
\citep{Safi22}, consistent with electrons of energy
$\sim 500\,\mathrm{TeV}$ emitting $\sim 4\,\mathrm{keV}$ photons
\citep{Hess2024,LHAASO}.

\subsection{Simulation targets}
The multi-wavelength picture described above identifies two concrete
observational constraints we include in our large-scale simulation of SS\,433:

\begin{enumerate}
  \item Quasi-stationary recollimation shocks at $\sim 30\,\mathrm{pc}$.
    The X-ray and TeV emission is concentrated in compact, standing knots
    at projected distances of $\sim 25$--$30\,\mathrm{pc}$ from the binary.
    Our simulations should form such structures naturally through the
    pressure imbalance between the jet and its surroundings.

  \item High, ordered X-ray polarization with the magnetic field
    aligned along the jet.
    The IXPE polarization fractions ($\sim 60\%$) and the jet-parallel
    EVPAs place strong constraints on the magnetic topology within the
    recollimation region.
    We use synthetic polarization maps to discriminate between poloidal,
    toroidal, and helical field configurations.
\end{enumerate}

We focus exclusively on the eastern lobe of SS\,433, where the IXPE
detection is most significant and the multi-wavelength data are most
complete.
As discussed in Sect.~\ref{sec:style}, our simulation domain spans the
region from $\sim 10\,\mathrm{pc}$ to $\sim 120\,\mathrm{pc}$ along the
jet axis, deliberately chosen to encompass the observed recollimation zone
while remaining computationally tractable.
Precession of the inner jet is not included in our setup.
At the distances we model ($\gtrsim 10\,\mathrm{pc}$), the
time-averaged precession cone is well approximated by a continuous,
axisymmetric, hollow outflow with a half-opening angle of $\simeq 6-10^{\circ}$
\citep{Fongantini,MonceauBaroux15, Goodall} or even $20^{\circ}$~\citep{Roberts}, and our fixed-direction
injection boundary captures this limiting behavior.
Jet--shell interaction and the detailed hydrodynamics of the W50 nebula
are not modeled; these processes are expected to become significant
further downstream and will be addressed in future work.

\section{Methods} \label{sec:style}
The simulations are performed in a Cartesian computational domain spanning $[-15,15]\times[0,40]\times[-15,15]$ in $(x,y,z)$ (see Table~1). Outflow boundary conditions are applied on all domain boundaries except at the lower boundary ($y=0$), where a reflecting boundary is used outside the jet nozzle and the jet material is continuously injected through a circular nozzle of radius $r_{\rm jet}$.

The ambient medium is initialized as homogeneous (see Tab.~\ref{tab:beta}). For models including a poloidal magnetic-field component (see, e.g., Sec.~\ref{subsec:mag}, the ambient medium is filled with the same uniform poloidal field as the injected jet, ensuring continuity of the magnetic field along the jet direction throughout the computational domain. The toroidal magnetic-field component is confined to and continuously injected through the jet nozzle with radius $r$, where the prescribed poloidal, toroidal, or helical magnetic-field configuration is maintained throughout the simulation.
See the following Sec.~\ref{subsec:mag} for a quantitative explanation of the initialized magnetic field components.


The jet injection nozzle has a radius of $r_\mathrm{jet} = 0.5$--$1.0$
in code units, corresponding to $5$--$10\,\mathrm{pc}$ in physical units
with our scaling $L_0 = 3.1 \times 10^{19}\,\mathrm{cm} \approx 10\,\mathrm{pc}$.
This scale represents the cross-sectional size of the \emph{already-established}
large-scale jet at the inner boundary of our computational domain, not at
the binary itself.
Injecting a jet at a downstream boundary with parameters appropriate for
its state at that location is standard practice in large-scale jet
simulations \citep[e.g.,][]{Fromm2016,delaCita,BowlerKeppens18,Kramer&MacDonald}, as we focus on formation, propagation, termination of the global jet structure. It is noteworthy that we understand that long-term localized jet physics might are treated in a different approach of a rather steady-state jet~\citep[see, e.g.,][]{Ricci}.
The physical motivation for this choice follows directly from the
geometry of SS\,433: \citet{MonceauBaroux15} show that the precessing
inner jet transitions to a time-averaged, approximately axisymmetric
hollow cone with a half-opening angle of $\theta \approx 10^{\circ}$ at
distances of several parsecs from the binary.
At our inner boundary ($\sim 10\,\mathrm{pc}$), a $10^{\circ}$
half-opening angle implies a jet cross-section of radius
$r \approx 10\,\mathrm{pc} \times \tan(10^{\circ}) \approx 1.8\,\mathrm{pc}$
for the hollow wall, but the full cone spans a diameter of
$\sim 2 \times 10\,\mathrm{pc} \times \sin(10^{\circ}) \approx 3.5\,\mathrm{pc}$,
consistent with our adopted injection radius of $5$--$10\,\mathrm{pc}$
when the finite width of the precession-smeared cone is accounted for.
The injection thus captures the physical extent of the time-averaged
outflow, not an arbitrary geometric parameter.


The downstream portion of the grid, $y \in [20, 40]$ code units
($200$--$400\,\mathrm{pc}$), lies beyond the region of physical interest
analyzed in this work and serves as an outflow buffer: it suppresses
spurious reflections from the downstream boundary that could otherwise
contaminate the recollimation zone at $y \lesssim 15$ code units.


\begin{table}
  \caption{Plasma beta of the ambient (external) medium for each
    magnetic-field model.
    $B_\mathrm{amb}$ is the ambient magnetic field strength in code
    units; $p_\mathrm{amb} = 10^{-5}$ in code units for all models;
    $\beta_\mathrm{amb} = 2 p_\mathrm{amb} / B_\mathrm{amb}^2$.}
  \label{tab:beta}
  \centering
  \begin{tabular}{llll}
    \hline\hline
    Model & $B_\mathrm{amb}$ & $\beta_\mathrm{amb}^1$ & Regime \\[0.08cm]
    \hline
    Pure poloidal        & $B_{z,0}$     & $4$--$8$  & Thermal \\
   Pure toroidal & 	$B_{z,0}$ & 	$4$–$8$ & 	Thermal          \\
    Helical 0.8/0.2 pol  & $0.8\,B_{z,0}$ & $6$--$12$ & Thermal \\
    Helical 0.6/0.4 pol  & $0.6\,B_{z,0}$ & $10$--$20$ & Thermal \\
    Helical 0.4/0.6 pol  & $0.4\,B_{z,0}$ & $25$--$40$ & Weakly magnetized   \\
    Helical 0.2/0.8 pol  & $0.2\,B_{z,0}$ & $100$     & Weakly magnetized   \\
    \hline
  \end{tabular}
  \tablefoot{For models with a non-zero poloidal component the ambient
    medium carries the same uniform poloidal field $B_{z,0}$ as the
    injected jet.
    In all thermally dominated cases ($\beta_\mathrm{amb} \gtrsim 4$)
    the ambient medium is consistent with estimates for the warm,
    moderately magnetized interior of the W50 nebula at distances of
    $\sim 30\,\mathrm{pc}$ from the binary \citep{Safi22}.
    The toroidal-only model has no ambient field component aligned with
    the jet axis, giving $\beta_\mathrm{amb} \to \infty$; a weak
    disordered field may be present in the W50 interior but is not
    modeled here.\\
    $^1$approximate values.}
\end{table}

\subsection{Magnetic field morphology}\label{subsec:mag}
The magnetic field strength $B$ can be defined in terms of geometry as:
\begin{equation}
    B_x = \frac{\frac{( y - y_c )}{a} \cdot b_m}{1 + ( \frac{r}{a} )^2},
    \quad
    B_y = - \frac{\frac{( x - x_c )}{a} \cdot b_m}{1 + ( \frac{r}{a})^2},
\end{equation}
where \((x_c, y_c)\) is the location of the center of the jet, which in our simulation is set to (0, 0). The variable \( r \) defines the jet’s radius while \( a=0.5 \) represents a magnetization radius. The parametrization constant of the magnetic field \( b_m \) is given by:

\begin{equation}
    b_m = \sqrt{\frac{-4p_j \sigma_\phi}{a^2( 2\sigma_\phi - 1 + 4 \log(a))}},
\end{equation}
where \( \sigma_\phi \) is the magnetization parameter for the toroidal component. The constant poloidal term \( B_z \) threading the jet is written as:

\begin{equation}
    B_z = \sqrt{\sigma_z ( b_m^2 a^2 + 2p_j )},
\end{equation}
where \( \sigma_z \) is the magnetization parameter of the poloidal component. We choose \( \sigma_z = 1 \) for a purely poloidal magnetic field and \( \sigma_\phi = 1 \) for a purely toroidal field~\citep{Nishikawa2019, Mignone2019, Kramer&MacDonald}. A helical field is produced by combining $\sigma_\phi \, \text{and} \, \sigma_z$. To scale the dimensionless grid units into physical units, the following conversions were applied:
\begin{equation*}
\rho_{\text{cgs}} = \rho \cdot \rho_0, \quad p_{\text{cgs}} = p \cdot \rho_0 v_0^2, \quad B_{\text{cgs}} = B \cdot \sqrt{4\pi \rho_0 v_0^2}
\end{equation*}
In terms of simulation parameters, the jet density varies from $\rho_{\text{jet}} = 0.001$ to $0.004$, contrasted against an ambient density of $\rho_{\text{ambient}} = 10.0$. Further, the unit values and fiducial parameters can be found in Tab.~\ref{tab:1}. 

\begin{table}[h!]
\centering
\begin{tabular}{|l|c|c|}
\hline
\textbf{Parameter}                            & \textbf{Parameter}           & \textbf{Value}   \\ \hline
Grid size                                     & $N \times N \times N$        & 300$\times$400$\times$300                 \\ \hline
Cell size                                     & $\Delta x \times \Delta y \times \Delta z$  &  30 $\times$ 40 $\times$ 30             \\ \hline
Unit density                                  & $\rho_0$                     &  $2.58\times 10^{-22}$             \\ \hline
Unit length                                   & $L_0$                        &   $3.1\times 10^{19}$               \\ \hline
Unit velocity                                 & $v_0$                        &   $2.99 \times 10^{10}$               \\ \hline
Bulk Lorentz factor                           & $\gamma$                     &    1.036              \\ \hline
Mach number                                   & $M$                          &     9             \\ \hline        
Pressure - Ambient               & $p_{\text{ambient}}$         &    $1.0\times 10^{-5}$              \\ \hline
Density - Ambient                & $\rho_{\text{ambient}}$      &   10.0               \\ \hline        

\end{tabular}
\caption{
Simulation parameters and unit conversions adopted in the RMHD models. The table lists the computational grid, physical scaling factors used to convert code units into cgs units, and the fiducial jet and ambient medium parameters common to all simulations. The last four parameters, pressure \& density for ambient medium and jet, respectively, are in code units.
}\label{tab:1}
\end{table}

\subsection{Relativistic magnetohydrodynamics}
The \texttt{PLUTO} code~\citep{Mignone2007} was used to model the dynamics of magnetized fluid flows. The governing system of conservation laws in RMHD is expressed as:

\begin{equation}
\partial_t U_k + \sum_{i \in \{x,y,z\}} \partial_i T^{ik} = 0 ~~.
\end{equation}

Here, \(U_k\) is the state vector of conservative quantities such as density (\(D\)), momentum (\(m_x, m_y, m_z\)), magnetic field components (\(B_x, B_y, B_z\)), and energy (\(E\)). The tensor \(T^{ik}\) represents the fluxes. For an ideal relativistic magnetized fluid, the conserved quantities and fluxes are:

\begin{equation}
U_k = \begin{pmatrix}
D \\
m_x \\
m_y \\
m_z \\
B_x \\
B_y \\
B_z \\
E
\end{pmatrix}, \quad T^{ik} = \begin{pmatrix}
D v_i \\
m_x v_i - b_x B_i \\
m_y v_i - b_y B_i \\
m_z v_i - b_z B_i \\
B_x v_i - v_x B_i \\
B_y v_i - v_y B_i \\
B_z v_i - v_z B_i \\
m_i
\end{pmatrix} ~~.
\end{equation}

The simulation assumes a specific equation of state (EoS) for a single-species relativistic gas, with the specific enthalpy \(h\) defined as:

\begin{equation*}
h = \frac{5}{2} \Theta + \sqrt{\frac{9}{4} \Theta + 1},
\end{equation*}
where \(\Theta = p/\rho\) is the ratio of pressure to density. Further, to control numerical violations of the solenoidal constraint ($\nabla\cdot\mathbf{B}=0$), we employ the eight-wave divergence-control formulation implemented in PLUTO~\citep{Powell94,Mignone2007}, which advects divergence errors with the flow while minimizing their dynamical effects.

\subsection{Synchrotron emission modeling}

Since MHD simulations do not directly constrain the non-thermal electron population, here we consider a power-law electron distribution of index $s=-2$ motivated by previous studies~\citep{Fromm2016, Kramer&MacDonald} as well as observational constraints on the photon index~\citep{Kaaret24}. The electron distribution is set by electron energy, $\gamma$, boundaries ranging from $\gamma_\text{min}=1$ to an exponential cutoff at $\gamma_{\rm max}$:
\begin{equation}
n_e(\gamma) = n_0 \gamma^{-2}e^{-\frac{\gamma}{\gamma_{\rm max}}}  ~~.
\end{equation}
Here, the normalization of the distribution, $n_0$, in every cell is proportional to the multiplication of kinetic energy and magnetic energy. However, if $n_0$ is chosen to be proportional to the kinetic energy or jet speed, the main conclusions remain the same (see Sec.~\ref{sec:polarization}. The normalization $n_0$ is then chosen to be representative for values to mimic the flux in the first recollimation shock is equal to the \textit{e1} knot in SS~433~\citep{Kaaret24}.

\subsection{Polarized radiative transfer}

The magnetic field and non-thermal particles derived from the above simulations are post-processed by the \texttt{3DPol} code~\citep{Zhang15}. This code calculates the Stokes parameters in each cell for the synchrotron emission, and ray-trace the emission to the plane of sky. Stokes parameters of the emission that arrives within the same time step and in the same cell on the plane of sky will be directly summed up, thus polarized emission from simulation cells with different magnetic field configurations can partially cancel the Stokes Q and U, reducing the observed polarization. The code includes all light crossing delays, although due to the large time step between snapshots, it is irrelevant to the present study.

\section{Magnetohydrodynamic results}\label{sec:mag}
We explore the dynamics of recollimation shocks in relativistic jets using high-resolution 3D simulations. Our simulations vary magnetic field configurations (poloidal, toroidal, and helical) and jet radii (in detail $r = 0.5$, $0.75$, $1.0$), and include multiple epochs to capture in-shock and off-shock states. The simulations are set up with uniform grids spanning $[-15, 15]$ in $x$ and $z$, and $[0, 40]$ in $y$, using resolutions up to $300 \times 400 \times 300$. We choose to zoom in to $x \in [-7,7]$ and $y\in [0,20]$. This paper will focus on the recollimation zone, the jet sheath-spine dynamic, as well as the influence of the underlying magnetic field on thermal properties as well as polarization. Polarimetric diagnostics include total intensity and linear polarization maps, EVPA distributions, and Stokes parameter ($I$, $Q$, $U$, $V$) profiles along the jet axis. 

For the following, Figs.~\ref{fig:pol}$-$\ref{fig:helall} are structured in a similar fashion. The top panels show pressure (color‑coded) and the bottom panels show density for radii 0.5 (left) and 0.75 (right); the case with radius $r=1$ is presented separately in Fig.~\ref{fig:helall}. The figures show 2D slices from 3D RMHD simulations of a jet launched from a nozzle with specified radius and varying initial conditions, forming a self‑consistent, collimated outflow with a bow shock propagating into a denser ambient medium and exhibiting Kelvin–Helmholtz instabilities in the backflow region.

The origin of the recollimation shocks can be understood from the
pressure balance between the jet and its confining cocoon.
In the toroidal and helical models with a dominant toroidal component,
the cocoon is mildly but distinctly over pressured with respect to the
ambient medium: the jet-to-cocoon pressure ratio at the first
recollimation point is $p_\mathrm{jet} / p_\mathrm{cocoon} \approx 0.5$--$0.7$,
meaning the cocoon pressure exceeds the local jet pressure by a factor
of $\sim 1.4$--$2.0$.
This lateral pressure excess drives the recollimation: the over pressured
cocoon compresses the jet inward, producing the oblique standing shock
at the radius where the jet pressure re-equilibrates.
The shock strength and standoff distance are in broad agreement with the
analytical recollimation model of \citet{DalyMarscher} for a cylindrical
jet in pressure imbalance, and with the numerical results of
\citet{Mizuno2015} for magnetized relativistic jets of comparable Mach
number.
In the purely poloidal model, by contrast, the jet-to-cocoon pressure
ratio at the recollimation point is $p_\mathrm{jet} / p_\mathrm{cocoon}
\approx 0.85$--$0.95$: the additional magnetic pressure of the poloidal
field partially resists lateral compression, resulting in weaker and more
widely spaced recollimation shocks.
This field-topology dependence of the cocoon overpressure and
recollimation shock strength is a key prediction of our simulations and
is directly testable through the spatial spacing of the observed X-ray
knots in SS\,433.

\subsection{Poloidal field configuration}
\begin{figure}
    \includegraphics[scale=0.4]{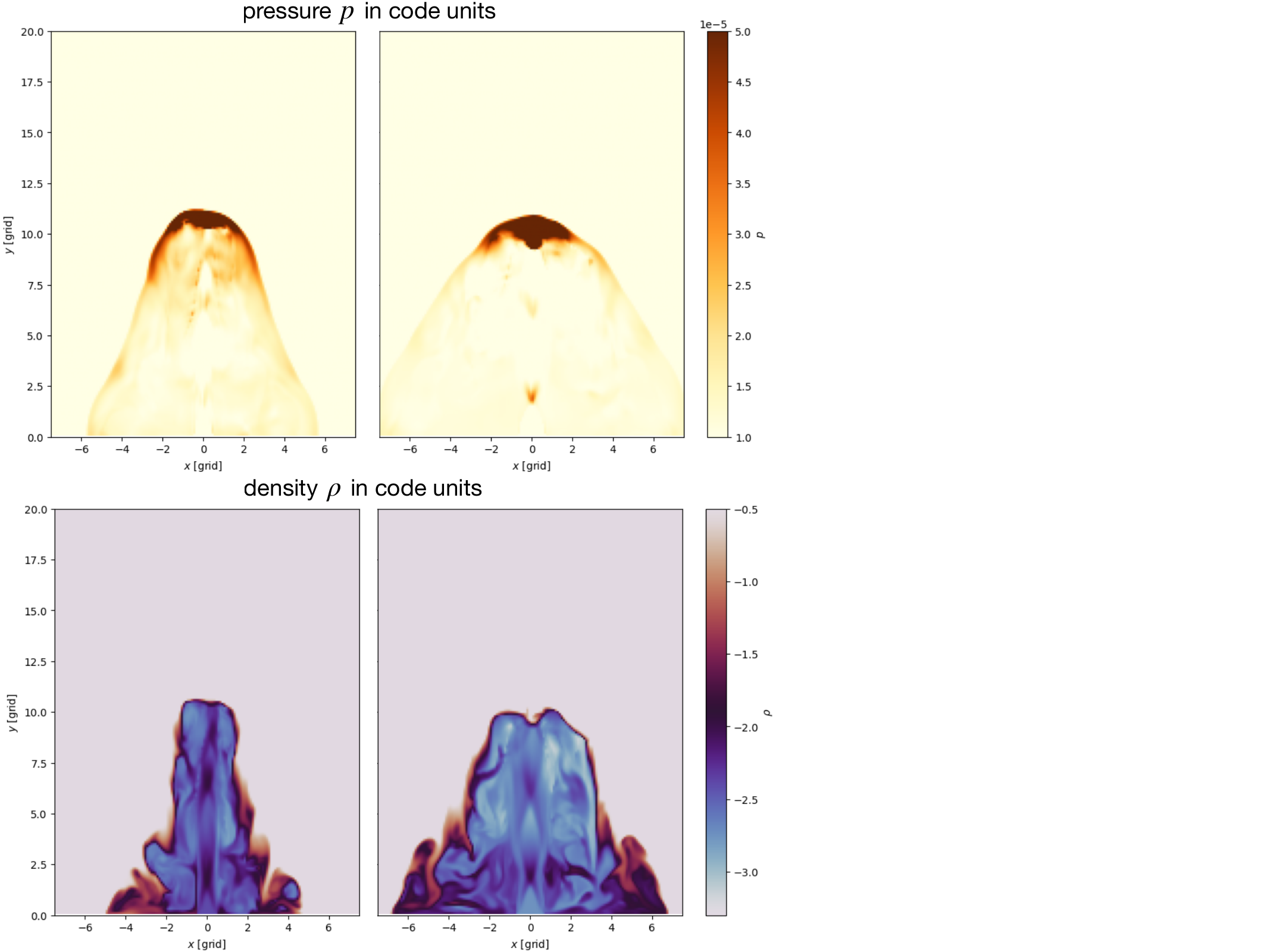}
    \caption{Pressure (top) and density (bottom) slices through the mid-plane of the 3D RMHD simulations for jets with a purely poloidal magnetic field and injection radii of $r=0.5$ (left) and $r=0.75$ (right).}
    \label{fig:pol}
\end{figure}

Jets launched with a purely poloidal field ($\vec{B} \parallel \vec{v_\text{jet}}$) show the least collimation (see~\ref{fig:pol}). The flow appears generally wider, and the shock front is flatter and broader compared with the toroidal case. The internal shock pattern is less regular, and recollimation features are weaker and more distributed.

A striking difference is visible in the intensity of the magnetic field at the primary recollimation shock: the field is strongly amplified there, resulting in a more magnetically dominated, high‑pressure zone directly behind the jet head (see top panel Fig.~\ref{fig:pol} and Fig.~\ref{fig:diag}). This is consistent with compression of the initially well structured field component. However, the density structures (bottom panel Fig.~\ref{fig:pol}) appear much smoother than in the toroidal case. Internal shocks are weak and spaced far apart, suggesting quasi-steady propagation.

\subsection{Toroidal field configuration}
\begin{figure}
    \centering
    \includegraphics[scale=0.4]{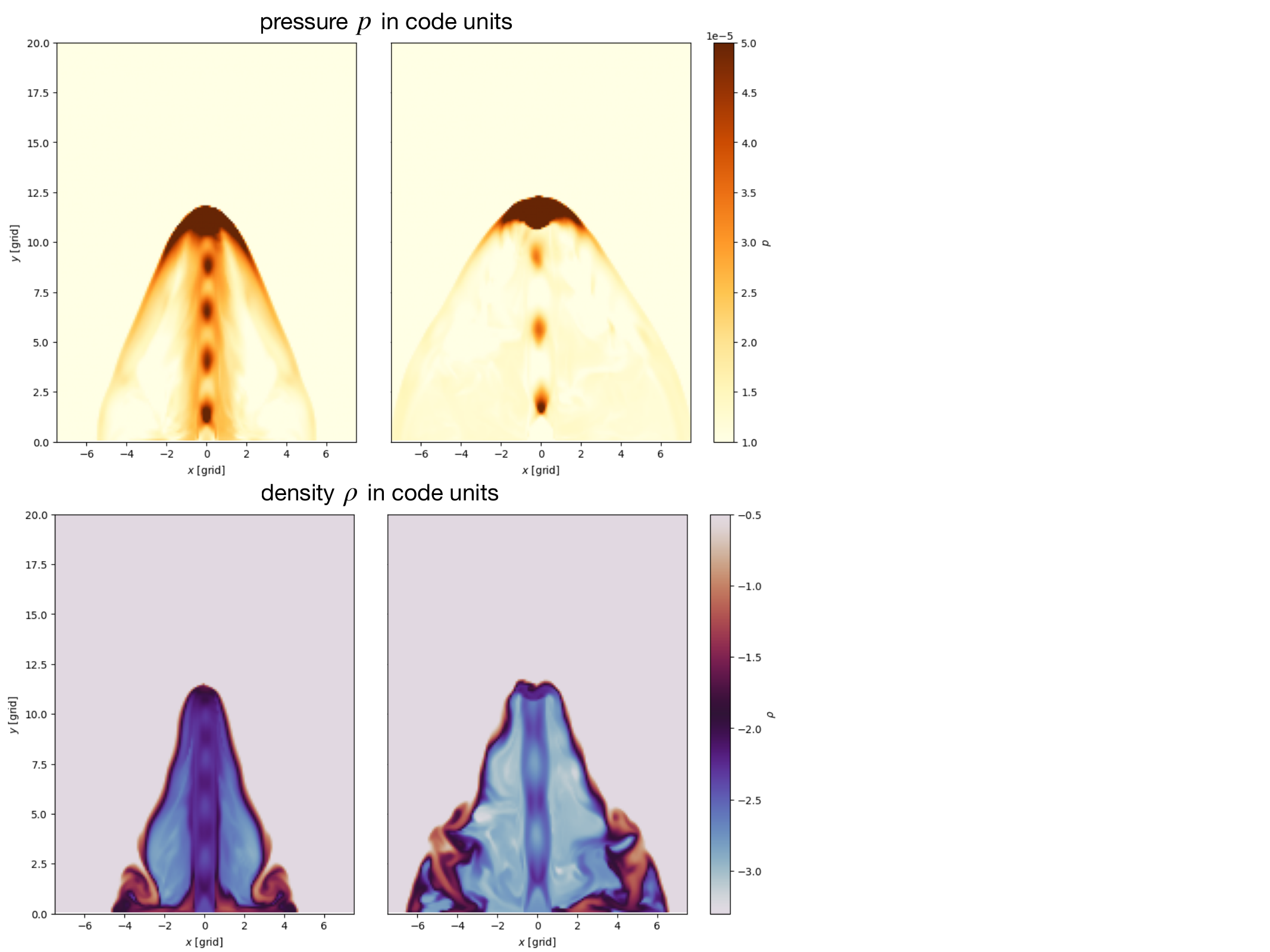}
    \caption{Same as Fig.~\ref{fig:pol}, but for a purely toroidal magnetic field.}
    \label{fig:tor}
\end{figure}

The toroidal models ($\vec{B} \perp \vec{v_\text{jet}}$) consistently produce the most tightly collimated jets among all configurations. Across both radii r = 0.5 and r = 0.75 (see Fig.~\ref{fig:tor}), the pressure maps (top panel Fig.~\ref{fig:tor}) show a narrow, well‑defined jet core, with a sequence of pronounced recollimation shocks. These shocks appear sharper and more frequent compared with the helical and poloidal cases (see for instance Fig.~\ref{fig:pol}.

The thermodynamic edge‑brightening visible in the density distribution (bottom panel~\ref{fig:tor}), while superficially similar across models, show a narrower spine and a more turbulent ambient medium as well as a more symmetric bow shock in the toroidal configuration.

The backflowing material contains no discernible (poloidal) component aligned with the flow, consistent with a rapid and turbulent field. This turbulence likely enhances the growth of Kelvin–Helmholtz instabilities, visible as small‑scale fluctuations in the density and pressure structure.

\subsection{Helical field configuration}
\begin{figure*}
    \centering
    \includegraphics[trim={0cm 8cm 0cm 0cm}, scale=0.5]{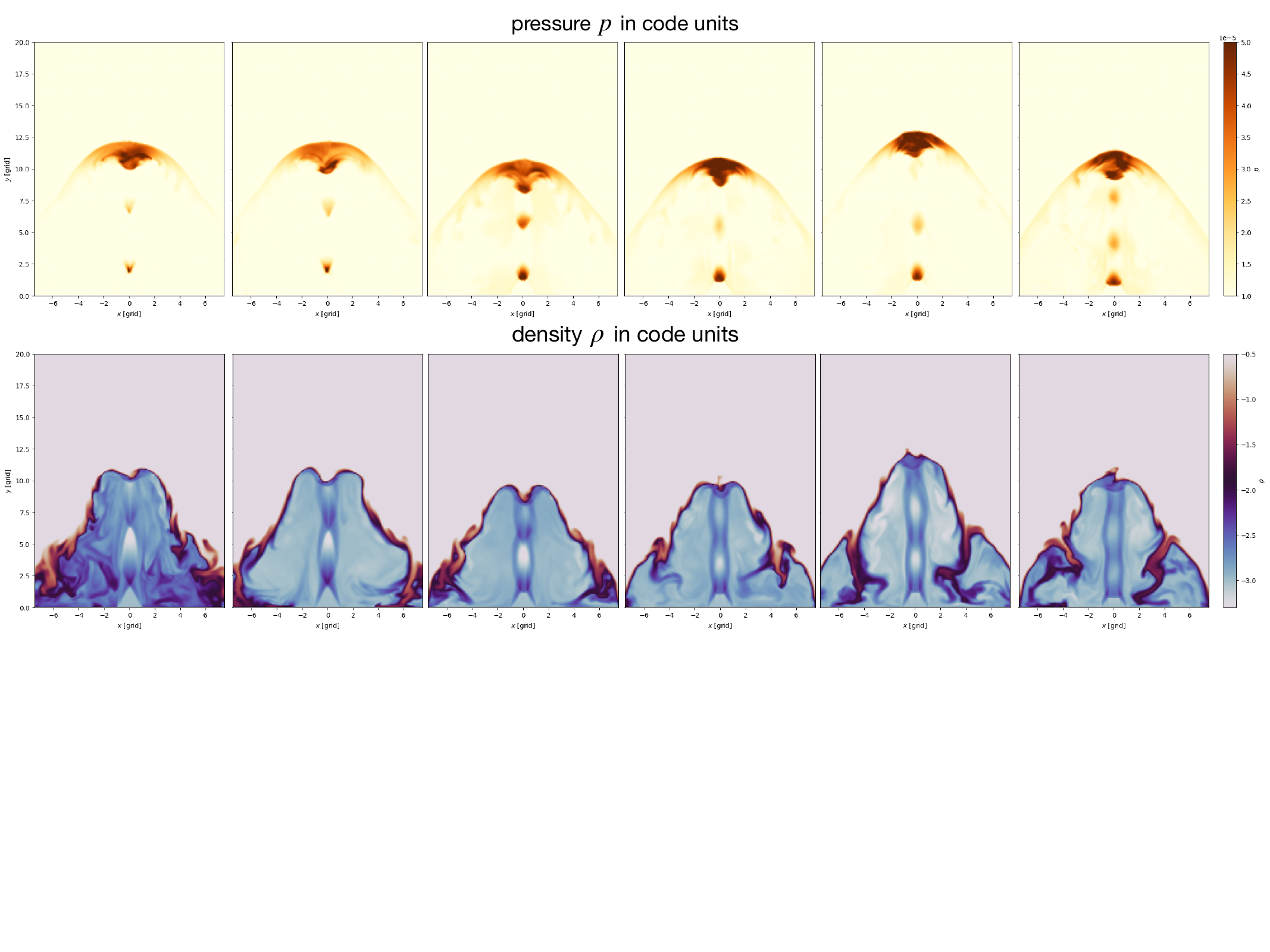}
    \caption{Pressure (top) and density (bottom) slices for jets with an injection radius of $r=1.0$ and different initial magnetic-field configurations. From left to right the magnetic topology changes from purely poloidal, through helical fields with poloidal-to-toroidal fractions of 0.8/0.2, 0.6/0.4, 0.4/0.6, and 0.2/0.8, to a purely toroidal field.}
    \label{fig:helall}
\end{figure*}

The intermediate helical models (with field‑component ratios, that is poloidal/toroidal, such as 0.8/0.2, 0.6/0.4, 0.4/0.6, 0.2/0.8) form a systematic sequence bridging the behavior of the poloidal and toroidal extremes. Across the sequence, the jet becomes progressively narrower and more structured as the toroidal fraction increases. Conversely, higher poloidal fractions promote broader jets with reduced shock contrast (see Fig.~\ref{fig:helall}).

Among the helical cases, the model with a dominant poloidal component (0.8/0.2) produces morphology closest to observations inferred from polarization analysis, providing a plausible best‑fit configuration (as discussed in Sec.~\ref{sec:polarization}). In this case, the jet retains moderate collimation while still forming well‑defined internal shocks, and the terminal shock remains compact but not overly constricted.

A consistent trend across the helical series is the evolution of the recollimation shock: from weak and relatively narrow in the poloidal‑dominated end to stronger, more pressurized, and spatially broader as the toroidal component is increased (see for instance top panel~\ref{fig:helall}). 

\subsection{Cross model trends}
\begin{figure}
    \centering
    \includegraphics[trim={18cm 0cm 0cm 0cm},scale=0.42]{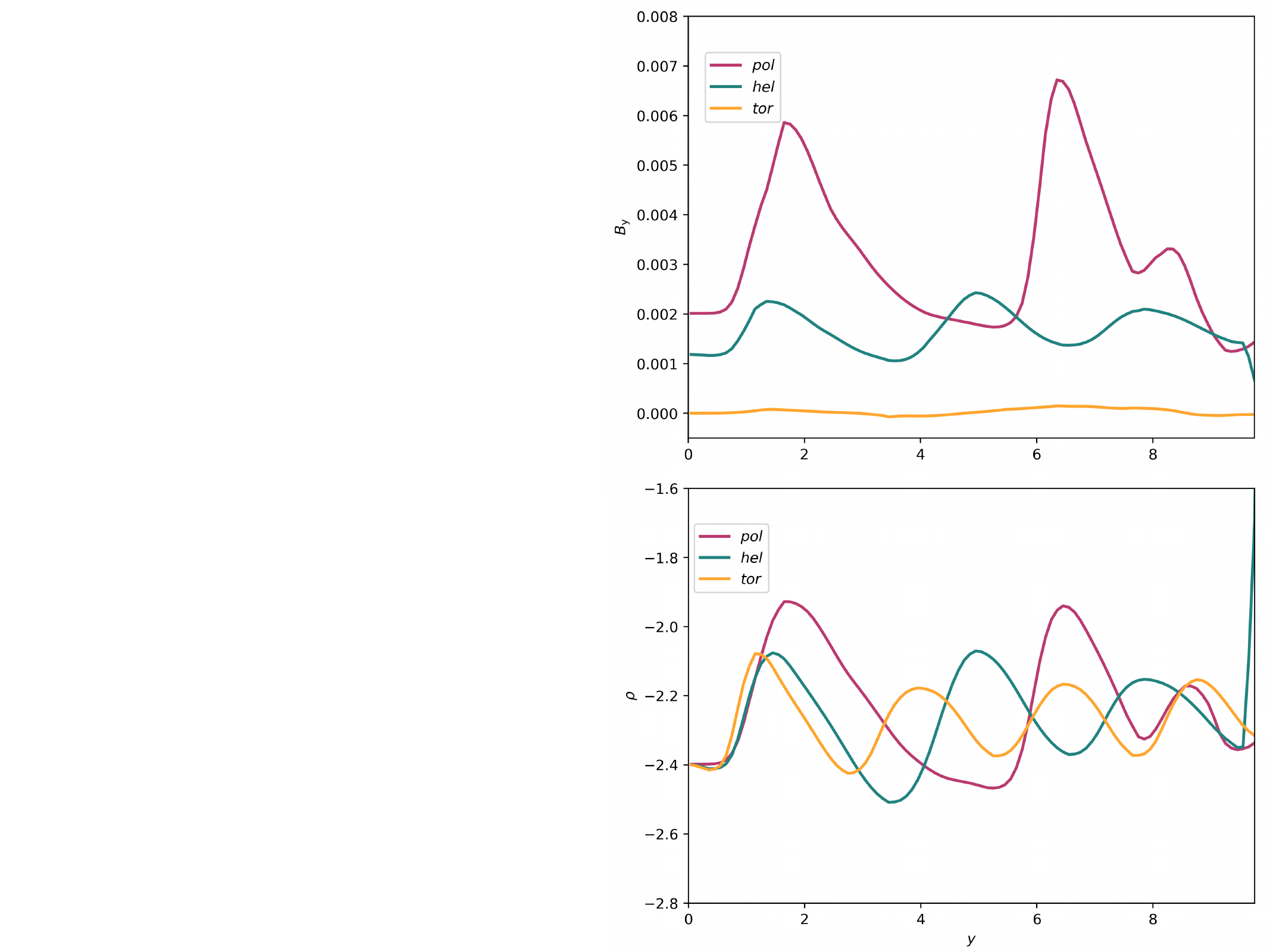}
    \caption{One-dimensional profiles extracted along the jet axis comparing the different magnetic-field configurations. The \textbf{upper} panel shows the magnetic-field strength in direction of the jet, while the \textbf{lower} panel presents the corresponding density distribution.}
    \label{fig:diag}
\end{figure}

Beyond the stated points (see previous sections), several additional trends are evident:

\begin{itemize}
    \item The bow shock morphology varies systematically: the toroidal models show a more pointed and elongated bow shock, while the poloidal models yield a rounder and flatter structure at the jet head.
    \item The cocoon region differs not only in turbulence but also in symmetry: toroidal runs show more symmetric side‑lobes and bow shocks, whereas poloidal and mixed‑field cases show asymmetric shear layers, likely tied to differing Kelvin-Helmholtz mode growth.
    \item The internal spine visible in the density maps fades substantially in the poloidal configuration but remains clear in toroidal and most helical cases.
    For larger nozzle radii (moving from $r = 0.5$ to $r = 0.75$), all field configurations show slightly wider jets and reduced shock contrast, but the qualitative differences between magnetic morphologues persist.
    \item The pressure contrast between the jet core and cocoon is maximal in the toroidal case, minimal in the poloidal case, and intermediate (and smoothly varying) in the helical morphologies.
    \item The transverse density profiles (Fig.~\ref{fig:diag}) reveal that the amplitude and frequency of oscillations in $\rho$ increase from poloidal $>$ helical $>$ toroidal, establishing a thermodynamic hierarchy in shock development. The toroidal case shows frequent small-scale variations superposed on the main density front, while the helical case displays intermediate periodicity, and the poloidal jet shows two significant jumps associated with the recollimation shocks. 
\end{itemize}


The magnetic field profiles (Fig.~\ref{fig:diag}, top panel) and corresponding cuts of $B_y$ demonstrate that the axial field component systematically decreases from poloidal $>$ helical $>$ toroidal. In the poloidal configuration, $B_y$ reaches values up to $6 \times 10^{-3}$ and shows multiple, well-defined peaks coinciding with mild compressions along the jet. The helical case exhibits reduced amplitude ($\sim 2 \times 10^{-3}$) and smoother oscillations reflecting alternating pitch and periodic field reversals, while the toroidal case remains an order of magnitude weaker ($\lesssim 10^{-4}$) with only localized enhancements at the edges.
This hierarchy confirms a clear dominance of the magnetic field in the jet direction in shock regions, following poloidal $>$ helical $>$ toroidal. The presence of strong longitudinal fields in the poloidal setup helps stabilize the flow against kink-like deformations, whereas the toroidal case redistributes magnetic energy azimuthally, promoting compression and reconnection zones. The amplitude correlation between $B_y$ peaks and density enhancements indicates that magnetic tension and compression jointly regulate the jet’s internal shock pattern and emissivity profile.

Across all three configurations, shock activity, and thermodynamic variability outcomes are tightly governed by magnetic topology. That is, shock development is most frequent in the toroidal case, moderate in the helical, and least pronounced in the poloidal configuration. On the contrary, the magnetic-field dominance along the jet axis follows the inverse trend — strongest for poloidal, intermediate for helical, and weakest for toroidal — particularly evident within shock-compressed regions.
From an observational perspective, these differences predict systematic trends in polarization behavior. Poloidal fields yield high, stable polarization with EVPAs aligned to the jet; helical fields introduce EVPA swings and rotation-measure gradients; and toroidal fields generate low central polarization but strong edge signatures.
Thus, the comparison demonstrates that both the thermodynamic structure and the observable emission properties of relativistic jets are direct diagnostics of their magnetic-field configuration.

\begin{figure}
\centering
\includegraphics[width=0.99\linewidth]{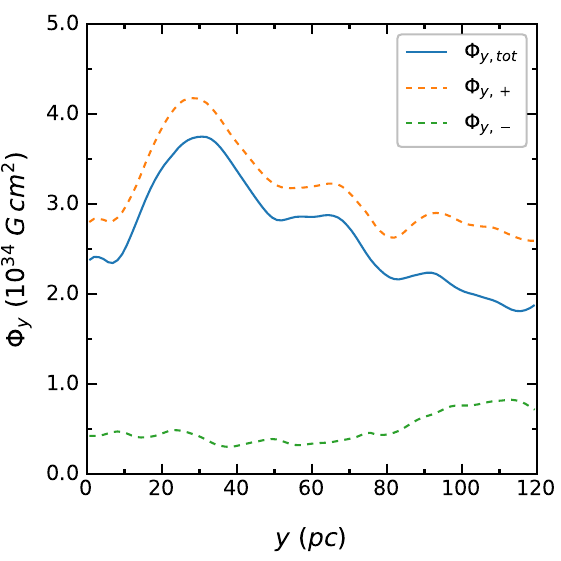}
\caption{The poloidal magnetic flux for the r=1.0 case with helical magnetic field with initially 0.8 in poloidal and 0.2 in toroidal components. Blue solid curve is the total magnetic flux, orange dashed curve is the positive magnetic flux, and green dotted curve is the absolute value of the negative magnetic flux.}
\label{fig:hel1ByFlux}
\end{figure}

\section{Magnetic flux and polarization structure of the jet: Results}\label{sec:polarization}

\begin{figure*}
\centering
\includegraphics[width=0.99\linewidth]{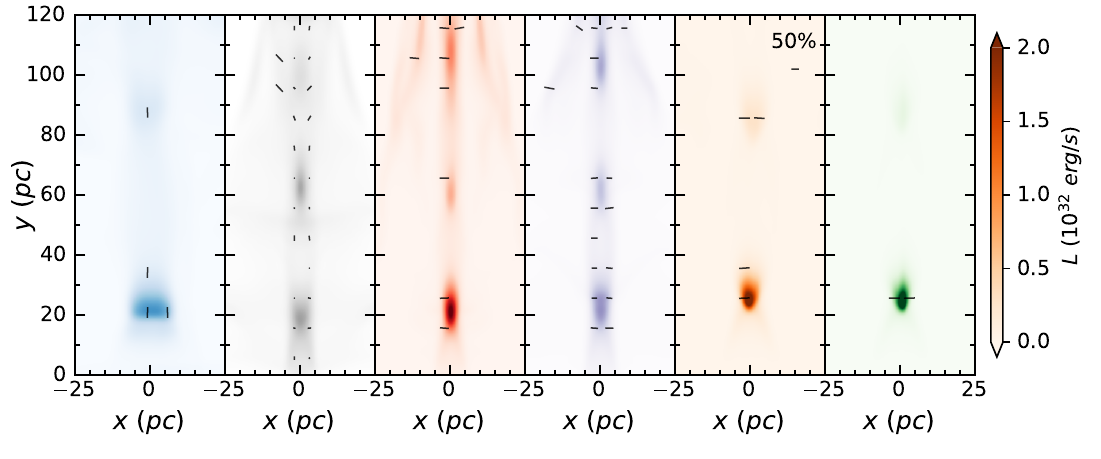}
\caption{Polarized emission maps for the r=1.0 cases with initially purely toroidal (blue), helical with 0.2 in poloidal and 0.8 in toroidal (grey), 0.4 in poloidal and 0.6 in toroidal (red), 0.6 in poloidal and 0.4 in toroidal (purple), 0.8 in poloidal and 0.2 in toroidal (orange), and purely poloidal (green) magnetic field. Color bar is the same for all cases, but only the orange is explicitly drawn. The length of the black dashes represents the polarization degree, and their direction is the electric vector position angle.}
\label{fig:torvshelvspol}
\end{figure*}

\begin{figure}
\centering
\includegraphics[width=0.99\linewidth]{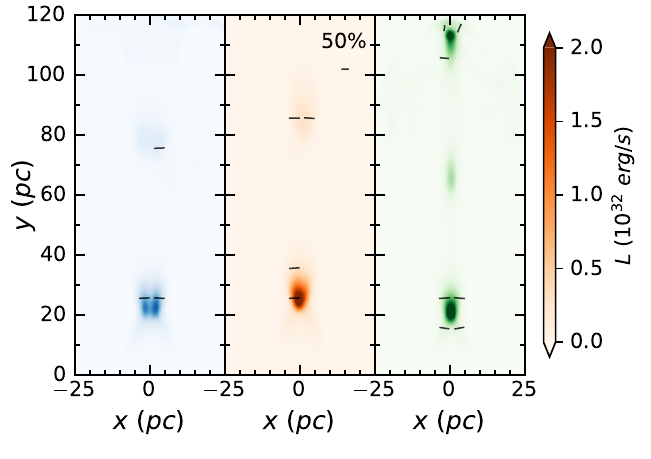}
\caption{Same as Figure \ref{fig:torvshelvspol}, but for r=0.5 (blue), r=1.0 (orange), and r=0.75 (green) cases with initially 0.8 in poloidal and 0.2 in toroidal case.}
\label{fig:hel05vs1vs075}
\end{figure}

\begin{figure}
\centering
\includegraphics[width=0.99\linewidth]{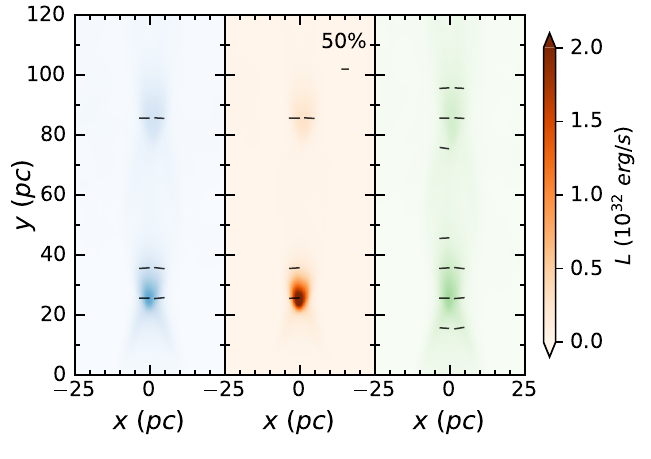}
\caption{Same as Figure \ref{fig:torvshelvspol}, but for different non-thermal electron normalization of the r=1.0 case with initially 0.8 in poloidal and 0.2 in toroidal case. Blue is normalized by kinetic energy, orange is normalized by the fiducial normalization by the multiplication of kinetic energy and magnetic energy, and green is normalized by the local jet speed.}
\label{fig:helkevskeb2vsv2}
\end{figure}

The evolution of the magnetic field and its imprint on the polarized synchrotron emission reveal how the interplay between poloidal and toroidal components shapes the structure of relativistic jets. Figures~\ref{fig:hel1ByFlux} $-$ \ref{fig:helkevskeb2vsv2} summarize these results for different magnetic configurations, jet radii, and electron energy normalizations.

Figure~\ref{fig:hel1ByFlux} presents the variation of the poloidal magnetic flux along the jet axis for an initially helical configuration with 80\% poloidal and 20\% toroidal field. The total flux $\Phi_\text{y, tot}$ (blue solid line in Fig.~\ref{fig:hel1ByFlux}) remains dominated by the positive magnetic component ($\Phi_\text{y,+}$, orange dashed line), while the absolute value of the negative contribution ($\Phi_\text{y,-}$, green dotted line) stays significantly smaller. The modest variations along y ($\vec{v}_\text{jet}$) and the small negative flux suggests some local field reversals or turbulent regions but not enough to cancel the overall poloidal field direction. The field remains largely ordered and aligned with the jet — consistent with a strong poloidal magnetic component.

The resulting polarized synchrotron emission, shown in Fig.~\ref{fig:torvshelvspol}, depends strongly on the initial magnetic geometry. The color scale represents total intensity, while the black dashes denote polarization vectors, whose lengths indicate the degree of polarization. For jets dominated by a toroidal field (first panel on the left), the polarization angle is generally aligned with the jet propagation direction. When the poloidal fraction increases to 40\%, the polarization angle becomes perpendicular to the jet direction. Importantly, the polarization angle in every recollimation shock is the same, and the polarization degree is comparable as well.

Figure~\ref{fig:hel05vs1vs075} extends this comparison to different jet radii ($r \in [0.5,0.75,1.0]$) for the same initial field with 80\% poloidal and 20\% toroidal components. Despite the changes in physical scale, the overall emission and polarization morphologies remain similar. This approximate self-similarity suggests that the large-scale polarization characteristics are primarily determined by magnetic structure rather than jet size.

In Fig.~\ref{fig:helkevskeb2vsv2}, the same magnetic field configuration is examined under different non-thermal electron normalizations. Here, the emissivity is normalized respectively by kinetic energy, by the product of kinetic and magnetic energy, and by kinetic energy with the local jet velocity. Nevertheless, the general polarized emission map appears similar, indicating that the exact non-thermal electron normalization does not affect the polarization signatures governed by the magnetic morphology.

\section{Discussion}\label{sec:disc}
\subsection{Recollimation shocks and magnetic regulation of the jet}
Our high-resolution three-dimensional RMHD simulations provide a framework that complements both the recent observational picture of SS\,433 presented by \citet{Hess2024} and theoretical studies of magnetized relativistic jets~\citep{Churazov}. The connection between internal shock formation, magnetic-field topology, and polarization structure offers new insight into the physical mechanisms governing jet stability and non-thermal particle acceleration in microquasar systems.

The recent H.E.S.S. observations identify a stationary, sub-relativistic shock in the eastern lobe of SS\,433 as the principal site of particle acceleration up to hundreds of TeV~\citep{Hess2024}. Our simulations naturally reproduce such quasi-stationary recollimation shocks through pressure imbalances between the jet and its surrounding medium. The morphology and strength of these shocks depend strongly on the magnetic topology.

Toroidal magnetic fields produce the strongest magnetic confinement, leading to narrow jets and a sequence of pronounced recollimation shocks. Such repeated compressions provide favorable conditions for efficient particle acceleration and are consistent with the compact TeV-emitting regions inferred from observations. In contrast, poloidal fields generate broader jets with weaker but more extended shocks, reflecting reduced magnetic confinement and smoother pressure equilibration. Helical fields bridge these two limiting cases, demonstrating that the balance between the poloidal and toroidal components regulates both the internal shock structure and the global jet morphology.

These results suggest that the magnetic field is not merely a passive tracer of the flow but actively governs the conversion of bulk kinetic energy into localized shock structures where high-energy particles are accelerated.

\subsection{Polarization constraints on the magnetic field geometry}
The synthetic polarization maps provide a direct diagnostic of the underlying magnetic-field topology. IXPE measured polarization fractions of $62\%\pm15\%$ in the eastern lobe and $59\%\pm17\%$ in the C11 region together with an approximately constant EVPA of $-12^\circ\pm10^\circ$, implying a highly ordered magnetic field aligned with the jet axis~\citep{Kaaret24,Hess2024}. Such observations place strong constraints on the magnetic geometry within the recollimation region.

Among the investigated models, the helical configuration with an initial poloidal-to-toroidal magnetic-energy ratio of approximately $0.8/0.2$ provides the closest agreement with the observations. Purely poloidal models reproduce the ordered polarization structure but under predict the degree of collimation and the strength of the recollimation shocks. Purely toroidal models, on the other hand, generate stronger magnetic confinement but also predict enhanced transverse EVPA variations, depolarization bands, and edge-brightened polarization structures that are not observed.

The preferred weakly helical model naturally combines both effects. The dominant poloidal field preserves the high polarization fraction and coherent EVPAs observed by IXPE, while the weaker toroidal component provides sufficient hoop stress to collimate the flow and sustain stationary recollimation shocks. The magnetic-flux evolution further supports this interpretation, showing that the positive poloidal flux remains dominant throughout the jet while local field reversals remain comparatively weak. Consequently, the large-scale magnetic ordering survives despite local turbulence generated within the shock region.

Our results therefore suggest that the recollimation region of SS\,433 is permeated by a predominantly poloidal magnetic field with a modest toroidal contribution. While the precise magnetic-energy ratio depends on the adopted model parameters, our simulations indicate that a weakly helical field containing roughly $80\%$ poloidal and $20\%$ toroidal magnetic energy best reproduces the currently available X-ray polarization observations.

\subsection{Comparison with previous magnetized jet simulations}
The magnetic-field dependence of recollimation shock properties found
here is consistent with and extends the results of several recent
numerical studies.
\citet{Mizuno2015} performed 3D RMHD simulations of recollimating
relativistic jets with toroidal, poloidal, and helical fields, finding
that toroidal and helical configurations enhance collimation and produce
stronger, more regularly spaced recollimation shocks, while purely
poloidal fields yield broader jets with weaker shocks — a trend our
simulations reproduce in the parameter space appropriate for SS\,433.
\citet{Moya21} studied the dynamics of magnetized relativistic jets
with helical field configurations in axisymmetric RMHD, finding that strongly
toroidal fields suppress the amplitude of radial jet oscillations through hoop
stress, yielding shallower pressure gradients and weaker recollimation shocks,
while more poloidal configurations allow larger lateral excursions and stronger
oblique shocks — a dynamical behavior consistent with the field-topology
dependence we find for the shock compression ratio across our helical model
sequence.
The companion Paper~II \citep{Fuentes2021} further showed that the synchrotron
emissivity peaks at the recollimation shock and depends sensitively on the
toroidal fraction; our synthetic polarization maps show an analogous behavior,
with the toroidal-dominated models producing compact, high-emissivity knots and
the poloidal models producing more extended, lower-contrast structures.
\citet{Boula26} performed 2D axisymmetric RMHD simulations of conical relativistic jets to quantify how magnetization $\sigma$, density contrast, pressure ratio, and magnetic pitch govern the geometry and strength of the first recollimation shock. They find that the recollimation distance decreases monotonically with increasing $\sigma$, scaling as $z_\mathrm{MHD}/z_\mathrm{HD} \propto (B_0^2/P_\mathrm{ext})^{1/3}$ in the magnetically dominated regime, and that the magnetic pitch is the primary factor controlling the local shock geometry: a dominant toroidal field yields compact, highly compressed recollimation zones with localized synchrotron emission knots, while a strong poloidal field shifts the recollimation point downstream and produces a more diffuse emission profile. These trends are fully consistent with the magnetic-topology dependence we find in 3D for the SS\,433 parameter regime: our toroidal models produce the most compact and luminous recollimation shocks, while the poloidal models produce broader, more extended structures, and the preferred 80\,\%/20\,\% poloidal/toroidal helical configuration occupies the intermediate regime that best matches the observed knot morphology.

\citet{Churazov} proposed an alternative origin for the EXJs in which
a collimated polar disk wind ($v \sim 2000$--$4000\,\mathrm{km\,s^{-1}}$,
half-opening angle $\sim 4$--$9^{\circ}$) is recollimated by the termination
shock of a surrounding isotropic outflow, rather than by the 0.26\,$c$
baryonic jets modeled here; both scenarios converge on quasi-stationary
oblique recollimation shocks as the particle acceleration sites.
Their model is purely hydrodynamical and does not address the magnetic field
structure, leaving the high polarization fraction ($\Pi \approx 60\,\%$;
\citealt{Kaaret24}) and the ordered field geometry detected by IXPE
unexplained — constraints that our RMHD simulations are specifically designed
to reproduce.

\subsection{Implications and limitations}

Our findings are consistent with recent theoretical work demonstrating that magnetic topology simultaneously regulates jet dynamics and observable polarization signatures on larger scales~\citep{Churazov}. By combining RMHD simulations with polarized radiative transfer, our study extends these ideas to the recollimation region of SS\,433 and establishes a direct connection between jet-scale MHD processes and X-ray polarization measurements.

The present simulations isolate the intrinsic magnetohydrodynamic evolution of the large-scale SS\,433 jet within a domain spanning $\sim 10$--$120\,\mathrm{pc}$ from the binary. Recollimation shocks arise self-consistently from the pressure imbalance between the overpressured cocoon and the jet: the jet-to-cocoon pressure ratio at the first recollimation point is $p_\mathrm{jet}/p_\mathrm{cocoon} \approx 0.5$--$0.7$ in the toroidal and helical models and $\approx 0.85$--$0.95$ in the poloidal model (Sect.~\ref{sec:mag}), reproducing the observed knot locations without any external driving. We do not, however, model the large-scale density and pressure gradients of the W50 interior, turbulence injected by the jet–nebula interaction, or a kinetic treatment of particle acceleration and transport. The ambient medium is initialized as a uniform, thermally dominated plasma with $\beta_\mathrm{amb} \approx 4$--$12$ (Tab.~\ref{tab:beta}), consistent with estimates for the W50 interior at $\sim 30\,\mathrm{pc}$ \citep{Safi22}, but any large-scale density gradient or pre-existing turbulence in the nebula is neglected. At larger distances from the binary, jet–shell interaction and ambient inhomogeneities are expected to become progressively more important, introducing additional depolarization and disrupting the ordered magnetic-field topology that characterizes the recollimation zone modeled here. A self-consistent treatment of these environmental processes, together with the non-thermal particle physics, is deferred to future work.

Our simulations adopt a fixed injection direction, treating the large-scale jet as a time-averaged, axisymmetric outflow. This approximation is well motivated at the distances we model: \citet{MonceauBaroux15} demonstrate that the precessing SS\,433 jet is dynamically recollimated into a continuous, effectively axisymmetric hollow cone already at ${\sim}0.07\,\mathrm{pc}$ from the binary
Future simulations incorporating a slowly precessing nozzle would be
needed to quantify this effect.

The physically analyzed portion of our simulation domain extends to
$\sim 20$ jet radii, which is sufficient to study
the first few recollimation shocks but is shorter than the $\gtrsim 100$
jet-radius domains used to study long-term jet propagation.
Whether instabilities
affect the X-ray and polarization properties of the SS\,433 knots over
the observed decade-long baselines remains an open question that our
current setup cannot address.
Longer-domain simulations with sufficient resolution to track these
instabilities are a natural next step.

\section{Conclusion}\label{sec:conc}
We have presented a suite of high-resolution three-dimensional RMHD simulations aimed at investigating the formation of recollimation shocks and their polarized synchrotron signatures in the jets of SS~433. By varying the magnetic-field topology between purely poloidal, purely toroidal, and mixed helical configurations, and by post-processing the simulations with polarized radiative-transfer calculations, we explored how the underlying magnetic structure regulates both the jet dynamics and the resulting observable emission.

Our simulations demonstrate that magnetic topology is the primary factor controlling the morphology and stability of the recollimation region. Toroidal-field jets produce the strongest confinement, the narrowest jet cores, and the most pronounced sequence of recollimation shocks. In contrast, poloidal-field jets exhibit broader outflows, weaker internal compressions, and smoother thermodynamic profiles. Helical configurations naturally bridge these two extremes, forming a continuous sequence of jet morphologies and shock strengths as the toroidal contribution increases.

The polarized emission maps reveal equally strong systematic trends. While the overall intensity distribution remains relatively insensitive to the assumed non-thermal particle normalization, the polarization structure is governed almost entirely by the magnetic geometry. Poloidal-dominated jets produce stable polarization patterns and EVPAs aligned with the jet axis, whereas increasing toroidal contributions lead to polarization rotations, enhanced transverse structure, and more complex magnetic-field signatures. Furthermore, the large-scale polarization morphology remains remarkably similar across different jet radii, suggesting that the magnetic topology rather than the physical jet size determines the observed polarization properties.

Comparison with recent IXPE observations of the eastern lobe of SS~433 favors a predominantly poloidal magnetic field within the recollimation region. In particular, the observed high polarization fractions and magnetic-field alignment with the jet direction are most naturally reproduced by the poloidal-dominated and weakly helical models. The toroidal configuration, while efficient at generating strong recollimation shocks, predicts polarization structures that are less consistent with the current X-ray observations.

For SS,433, the observational evidence identifies the quasi-stationary recollimation shocks — not the terminal interaction region — as the primary sites of non-thermal emission. Our results support a scenario in which the stationary X-ray and TeV emission regions of SS~433 are associated with magnetically regulated recollimation shocks that accelerate particles locally and maintain a highly ordered magnetic field. More broadly, this work demonstrates that polarization measurements provide a powerful probe of jet magnetic structure and offer a direct link between relativistic MHD processes and high-energy radiation. Future X-ray polarimetric observations with improved spatial resolution, combined with multi-wavelength studies and more realistic simulations including particle acceleration and jet–environment interactions, will further constrain the magnetic composition and dynamical evolution of relativistic jets in both Galactic microquasars and active galactic nuclei.

\begin{acknowledgements}
The authors thank Laura Olivera-Nieto for fruitful discussions on observations and interpretations of previous studies on SS~433.
JAK is supported by a European
Research Council Synergy Grant `BlackHolistic' grant No. 1010716. HZ is supported
by NASA under award number 80GSFC24M0006 and also
IXPE GO program Cycle 1, grant number 80NSSC24K1173. HL acknowledges the support by
the DOE FES program and NASA's IXPE program under 80NSSC26K0264. 

\end{acknowledgements}

\bibliographystyle{aa}
\bibliography{thesis_refs-2}

\begin{appendix}
    
\end{appendix}

\end{document}